\documentclass[aps,preprint,preprintnumbers,nofootinbib,showpacs]{revtex4}
\usepackage{graphicx,color,amsmath,amsfonts}

\newcommand{\beq}     {\begin{equation}}
\newcommand{\eeq}     {\end{equation}}
\newcommand{\bea}     {\begin{eqnarray}}
\newcommand{\eea}     {\end{eqnarray}}
\newcommand{\pmns}{\mbox{$ U_{\rm PMNS}$}}
\newcommand{\bad}{\begin{array}{ccc}}
\newcommand{\ea}{\end{array}}

\newcommand{\gauge}{SU(2)$_L\,\times\, $SU(2)$_R\,\times\,$U(1)$_{B-L}\,$}

\newcommand{\lsim}{\mathrel{\mathop{\kern 0pt \rlap
  {\raise.2ex\hbox{$<$}}}
  \lower.9ex\hbox{\kern-.190em $\sim$}}}
\newcommand{\gsim}{\mathrel{\mathop{\kern 0pt \rlap
  {\raise.2ex\hbox{$>$}}}
  \lower.9ex\hbox{\kern-.190em $\sim$}}}
\newcommand{\no}     {\nonumber}
\newcommand{\di}      { \mathrm{d }}

\newcommand{\lm}      {\lambda}
\newcommand{\es}      {\epsilon}

\newcommand{\Gm}      {\Gamma}

\newcommand{\dt}      {\delta}
\newcommand{\Dt}      {\Delta}
\newcommand{\kp}      {\kappa}

\newcommand{\gev}      {\;{\rm GeV}}
\newcommand{\tev}      {\;{\rm TeV}}

\newcommand{\afb}      {A_{\rm FB}}
\newcommand{\shat}      { \hat{s} }

\def\n{{(n)}}

\def\fan{{f_A^{(n)}}}

\newcommand{\Z}{\mathbb{Z}_2}

\newcommand{\szz}{S^1/\mathbb{Z}_2\times \mathbb{Z}_2'}

\begin{document}

\title{
Custodial bulk Randall-Sundrum model and $B \to K^* l^+ l^{\prime -}$
}

\author{Sanghyeon Chang}
\email{schang@cskim.yonsei.ac.kr}
\author{C.S. Kim}
\email{kim@cskim.yonsei.ac.kr}
\affiliation{
Department of Physics and IPAP, Yonsei University,
Seoul 120-749, Korea}
\author{Jeonghyeon Song}%
 \email{jhsong@konkuk.ac.kr}
\affiliation{%
Department of Physics, Konkuk University,
                   Seoul 143-701, Korea
}
\date{\today}

\begin{abstract}
\noindent The custodial Randall-Sundrum model
based on \gauge generates new flavor-changing-neutral-current (FCNC) phenomena
at tree level, mediated by Kaluza-Klein neutral gauge bosons.
Based on two natural assumptions of universal 5D Yukawa couplings and
no-cancellation in explaining the observed standard model fermion mixing matrices,
we determine the bulk Dirac mass parameters.
Phenomenological constraints from lepton-flavor-violations
are also used to specify the model.
From the comprehensive study of $B \to K^* l^+ l^{\prime -}$,
we found that only the
$B \to K^* e^+ e^-$ decay has sizable new physics effects.
The zero value position of the forward-backward asymmetry
in this model is also evaluated, with about 5\%
deviation from the SM result.
Other effective observables are also suggested such
as the ratio of two differential
(or partially integrated) decay rates
of $B \to K^* e^+ e^-$ and $B \to K^* \mu^+ \mu^-$.
For the first KK gauge boson mass of $M_A^{(1)}=2 -4 \tev$,
we can have about $10-20\%$ deviation from the SM results.
\end{abstract}

\maketitle

\section{Introduction}
\label{sec:introduction}

The standard model (SM) has been very successful in reproducing nearly all
experimental data on the fundamental interaction among gauge bosons and fermions.
Nevertheless, the SM is not regarded as a fully satisfactory theory
since it cannot explain two kinds of hierarchy:
One is the gauge hierarchy, and the other is
the hierarchy among the SM fermion masses.
In the SM, the hierarchies are attributed to the hierarchical parameters --
the bare Higgs mass parameter for the gauge hierarchy and
the Yukawa couplings for the fermion mass hierarchy.

Randall and Sundrum (RS) scenario with bulk fermion fields is one of
the rare candidates to be able to explain both
hierarchies\,\cite{Randall:1999ee,Goldberger:1999wh,Davoudiasl:1999tf,Chang:1999nh}.
The gauge hierarchy problem is explained by a geometrical
exponential factor.
Small SM fermion masses, which are proportional
to the overlapping probability of the bulk fermion wave function with
the confined Higgs boson field at the TeV brane, can be generated with
moderate values of the bulk Dirac mass parameters
\cite{Grossman:1999ra,
Gherghetta:2000qt,Huber:2003tu,Chang:2005ya,Agashe:2004ay,Agashe:2004cp}.
Yet the naive bulk RS model suffers from the strong constraints of
the electroweak precision data (EWPD): The first Kaluza-Klein (KK) mode
mass should be above $\sim 20\tev$
\cite{Kim:2002kk,Csaki:2002gy,Hewett:2002fe,Burdman:2002gr}. This is
due to the lack of SU(2) custodial symmetry. In
Ref.\,\cite{Agashe:2003zs}, an attractive model was proposed such
that the custodial  symmetry is induced from AdS$_5$/CFT feature
of bulk gauge symmetry of \gauge.

\begin{figure}[b!]
  \includegraphics[scale=1]{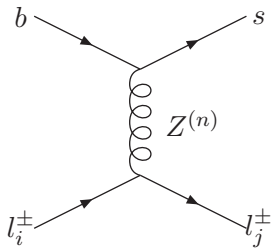}
   \caption{\label{fig:feyn}
Feynman diagram leading to $B \to K^* l_i^+ l_j^{-}$ mixing in a warped extra
dimension model. $Z^{(n)}$ is the $n$-th KK mode of the SM $Z$ boson.}
\end{figure}

One of the interesting features of this model is new
flavor-changing-neutral-current (FCNC) \emph{at tree level}, mediated by KK
gauge bosons\,\cite{FCNC:RS}. This is due to the misalignment between the gauge
eigenstates and the mass eigenstates: The five-dimensional
(5D) Yukawa interaction is
not generally flavor-diagonal. The fermion mass eigenstates of
different generations can couple with KK modes of a neutral gauge
boson, as one example is depicted in Fig.\,\ref{fig:feyn}. Note that
FCNC at tree level in this model involves four external fermions:
There is no tree-level effect on $b \to s \gamma$, for example. Since
FCNC in the SM occurs only at loop level, rare FCNC decays can be
a good place to probe the model.

Phenomenologically meaningful question is
whether we have reliable predictions for various FCNC processes.
The SM fermion mass matrix $M^f_{ij}$ answers,
which is determined by two ingredients.
One is the 5D Yukawa couplings $\lm^{f}_{5ij}$,
and the other is the fermion mode function
fixed by the bulk Dirac mass parameter $c_F$.
Unfortunately, there is no unique way to determine
both ingredients only from the observed SM fermion masses and mixing matrices,
albeit extensive studies presenting the feasibility of
the generation of SM fermion masses
by controlling the bulk Dirac mass parameters.

One reasonable approach is
to adopt minimal and natural assumptions.
In this paper, we have two basic assumptions.
The first one is that the 5D Yukawa couplings are universal,
{\it i.e.,}
$\lm_{5ij} \simeq \lm_5 \sim \mathcal{O}(1)$.
Small masses of the SM fermions
are explained by suppressed zero mode functions.
Second, we assume that when explaining the observed SM mixing matrices,
Cabibbo-Kobayashi-Maskawa (CKM) and Pontecorvo--Maki--Nakagawa--Sakata
(PMNS) matrices, each of which is
the product of two independent mixing matrices in this model,
no order-changing by cancellation is allowed.
Our choice has the least hierarchy,
which is consistent with the main motivation of this model.
Based on these two assumptions we examine
whether all the bulk Dirac mass parameters
as well as mixing matrices
can be fixed, and whether we have
reliable predictions for the phenomenological signatures of  FCNC process
such as $B \to K^* l^+ l^-$\,\cite{BKll:exp,BKll:Afb}.
This is our primal goal.

In the quark sector,
our two natural assumptions
are to be shown enough to fix all the bulk
Dirac mass parameters.
In the lepton sector,
there are some ambiguities due to
the observed \emph{large} mixing angles.
We will examine the constrains from lepton-flavor-violating processes
and determine the bulk lepton sector fairly accurately,
which is one of our new results.

With the phenomenologically specified parameters,
we will study the effect of the custodial bulk RS model
on various observables of $B \to K^* l^+ l^-$.
This decay mode, especially with $K^*$,
has several virtues in the experimental aspect.
As well as producing very clean signature,
its branching ratio is larger than the decay into $K$.
In addition, a vector boson $K^*$ decaying to $K\pi$ allows us
various angular analysis to measure many observables, such as
the forward-backward asymmetry $\afb$.
$\afb$ is a very good observable
to probe new physics effect,
since the so-called zero value position of $\afb(\shat_0)=0$
has strongly suppressed hadronic uncertainty
in the calculation of the form factors.
In addition, we present other sensitive probes of this model such as
the ratio of differential decay rates for $B \to K^* e^+ e^-$
and $B \to K^* \mu^+ \mu^-$.
The sensitivity is due to sizable coupling of $Z^{(1)}$-$e^+$-$e^-$
but suppressed coupling of $Z^{(1)}$-$\mu^+$-$\mu^-$
in this model.

The organization of the paper is as follows.  In Sec. \ref{sec:RS}, we
briefly review the custodial bulk RS model
with \gauge.
In Sec. \ref{sec:fermion:mass},
we formulate the bulk fermion sector, and
determine all the bulk Dirac mass parameters
based on our two natural assumptions.
We will show the inevitable ambiguity in the lepton sector
due to the large mixing angles.
Section \ref{sec:FCNC} deals with the FCNC in this model.
In Sec.\,\ref{sec:FV}, we examine the lepton-flavor violating processes,
and the new effect on
$B \to K^* l^+ l^-$.
Unique and sensitive observables to this model
are also proposed.
We conclude in Sec. \ref{sec:conclusions}.

\section{The Bulk RS Model: Basic formulae}
\label{sec:RS}

The RS model is based on
a 5D warped spacetime with the metric\;\cite{Randall:1999ee}
\begin{equation}
ds^2= e^{-2\sigma(y)}(dt^2-d\vec{x}^2) - dy^2,
\label{metric}
\end{equation}
where the fifth dimension $y \in [\, 0,L]$
is compactified on the $\szz$ orbifold, and the warped function is $\sigma(y) = k|y|$
with $k$ at the Planck scale $M_{\rm Pl}$.
There are two reflection
symmetries under $\Z:y \to -y$ and $\Z':y'(=y-L/2) \to -y'$.
Two boundaries are the $\Z$-fixed point at $y=0$
(Planck brane), and the $\Z'$-fixed point at $y=L$
(TeV brane).
In what follows, we denote $(\Z,\Z')$ parity
by $(\pm,\pm)$.
In many cases, conformal coordinate $ z\equiv e^{\sigma(y)}/k$ is more convenient:
\begin{equation}
ds^2= \frac{1}{(kz)^2}(dt^2-dx^2 - dz^2).
\end{equation}
With $kL \approx 35$, the natural cut-off of the theory $T\equiv e^{-kL}k$
becomes at the TeV scale, which answers the gauge hierarchy problem:
\begin{equation}
T \equiv\epsilon k \sim {\rm TeV} \quad
{\rm with}~ \epsilon \equiv e^{-kL} \sim \frac{\rm TeV}{M_{\rm Pl}}.
\end{equation}

We adopt the model suggested by  Agashe {\it et.al.} in Ref.\,\cite{Agashe:2003zs},
based on the gauge structure of
{SU(3)$_c\,\times\, $SU(2)$_L\,\times\, $SU(2)$_R\,\times\,$U(1)$_{B-L}\,$}:
The custodial symmetry is guaranteed by the bulk SU(2)$_R$ gauge symmetry.
The bulk gauge
 symmetry SU(2)$_R$ is broken into U(1)$_R$ by the orbifold boundary conditions on
the Planck brane
such that gauge fields
$\widetilde{W}_R^{1,2}$ have $(-+)$ parity.
The U(1)$_R\, \times\,$U(1)$_{B-L} $ is
spontaneously broken into U(1)$_Y$  on the Planck brane,
and the Higgs field localized on the TeV brane is responsible for the breaking of
SU(2)$_L\, \times\,$U(1)$_Y$ to U(1)$_{\rm EM}$.

A 5D gauge field
$A^M(x,z)$ is expanded in terms of KK modes,
\begin{eqnarray}
\label{eq:KKexpA} A_{\nu} (x,z) = \sqrt{k} \sum_n A_\nu^{(n)}(x)
\fan (z),
\end{eqnarray}
where the zero mode function is
$
 f_A^{(0)} ={1}/{\sqrt{k L}}
$.
The massless zero mode 
is interpreted as a SM gauge field\,\cite{Davoudiasl:1999tf}.
The general $\fan (z)$ function can be found in early references, for example, in
Ref.\,\cite{Chang:2005ya}. The bulk fermion field $\Psi(x,z) \equiv e^{2\sigma} \hat{\Psi}$
is also expanded as
\begin{equation}
\label{eq:KKexpansion:Psi}
\hat{\Psi}(x,z) =
\sqrt{k} \sum_{n} \left[ \psi^\n_L(x) f_{L}^\n(z) + \psi^\n_R(x)
f^\n_R(z) \right] \,.
\end{equation}
Two zero mode functions are
\begin{eqnarray}
\label{eq:fLR0}
 f_{L}^{(0)}(z,c) =f_{R}^{(0)}(z,-c)
 = \frac{(T z)^{-c}}{N_L^{(0)}}, 
\end{eqnarray}
where $c$ is defined by the bulk Dirac mass $m_D \!= \! c  k \ {\rm sign} (y)$,
and $N_L^{(0)}$ is referred to Ref.\,\cite{Chang:2005ya}.
Note that a massless SM fermion corresponds to the zero mode with $(++)$ parity.
Since $\Psi_L$ has always opposite parity of $\Psi_R$,
the left-handed SM fermion is the zero mode of a 5D
fermion whose left-handed part has $(++)$ parity. The
right-handed part has automatically $(--)$ parity which cannot describe a SM fermion.

This characteristic feature of a bulk fermion in a warped model
requires to extend the fermion sector.
For each left-handed SM fermion, there should exist another bulk fermion
whose right-handed part has $(++)$ parity.
Due to the gauge structure of \gauge,
these right-handed SM fermions belong to a SU(2)$_R$ doublet.
Since
$\widetilde{W}_R^{1,2}$ fields have $(-+)$ parity and couple two
elements of a SU(2)$_R$ doublet, one of the SU(2)$_R$ doublet should have
 $(-+)$ parity.
As a result, the whole quark sector is
\begin{equation} Q_i =\left(
       \begin{array}{c}
         u_{iL}^{(++)} \\
         d_{iL}^{(++)} \\
       \end{array}
     \right),
     \quad
U_i =\left(
       \begin{array}{c}
         u_{iR}^{(++)} \\
         D_{iR}^{(-+)} \\
       \end{array}
     \right),
     \quad
D_i =\left(
      \begin{array}{c}
        U_{iR}^{(-+)} \\
        d_{iR}^{(++)} \\
      \end{array}
    \right),
\end{equation}
and lepton sector is
\begin{equation} L_i =\left(
       \begin{array}{c}
         \nu_{iL}^{(++)} \\
         e_{iL}^{(++)} \\
       \end{array}
     \right),
     \quad
N_i =\left(
       \begin{array}{c}
         \nu_{iR}^{(++)} \\
         E_{iR}^{(-+)} \\
       \end{array}
     \right),
     \quad
E_i =\left(
      \begin{array}{c}
        N_{iR}^{(-+)} \\
        e_{iR}^{(++)} \\
      \end{array}
    \right),
\end{equation}
where $i$ is the generation index.
Dirac mass parameters ($c_{Q_i}$, $c_{U_i}$, $c_{D_i}$,
$c_{L_i}$, $c_{E_i}$, $c_{N_i}$)
determine the fermion
mode functions, KK mass spectra, and coupling strength with KK gauge bosons.

\section{The SM Fermion masses and mixings}
\label{sec:fermion:mass}

\subsection{Basic Assumptions}
\label{subsec:natural:assumption}

\begin{figure}
 \includegraphics[scale=0.65]{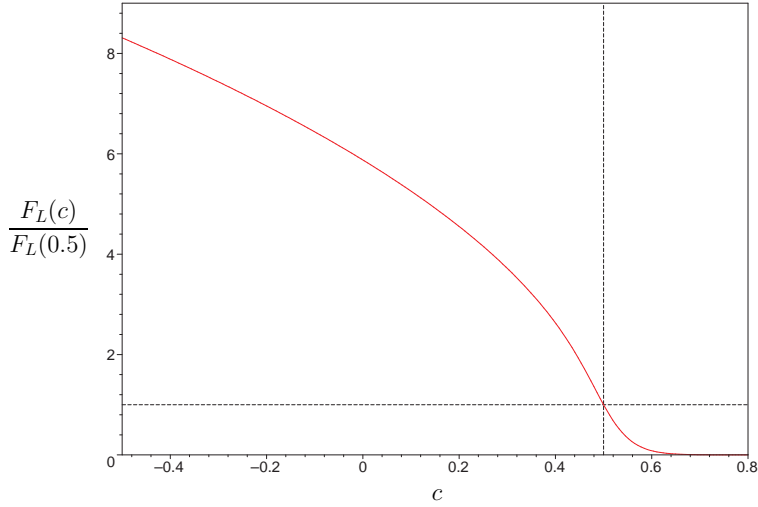}
\caption{$F_{L}(c) / F_L(0.5)$ as a function of the  bulk Dirac mass parameter $c$.}
\label{fig:FL}
\end{figure}

On the TeV brane, the SM fermion
mass is generated as the localized Higgs field develops its vacuum expectation value
of $\langle H \rangle = v\simeq 174\gev$.
The SM mass matrix for a fermion $f (= u,d,\nu,e)$ is
\begin{equation}
\label{eq:Mf:general}
\big(M_f \big)_{ij}= v \lambda^f_{5ij}
\left. \frac{k}{T} \, f_R^{(0)}(z,c_{R_i}) f_L^{(0)}(z,c_{L_j})
\right|_{z=1/T}
\equiv v \lambda^f_{5ij} F_R(c_{R_i}) F_L(c_{L_j}),
\end{equation}
where $i,j$ are the generation indices,
$\lambda^f_{5ij}$ are the 5D (dimensionless) Yukawa couplings
and 
\beq
F_{L}(c)=F_R(-c) \equiv \frac{ f_{L}^{(0)}\left( {1}/{T},c \right) }{\es^{1/2}}
.
\eeq
Figure \ref{fig:FL} shows $F_{L}(c)$, normalized by $F_{L}(0.5)$,
as a function of $c$.
The value of $F_L(c)$ decreases with increasing $c$,
and becomes suppressed
once $c>0.5$.

The mass eigenstates of the SM fermions
are then
\begin{equation}
\label{eq:chi-psi-mix}
\chi_{fL}=U^\dagger_{fL} \psi^{(0)}_{fL},\quad \chi_{fR}=U^\dagger_{fR} \psi^{(0)}_{fR}.
\end{equation}
Note that the observed mixing matrix is a multiplication of \emph{two
independent} mixing matrices
such that $V_{\rm CKM}  = U_{uL}^\dagger U_{dL}$
and $\pmns = U_{e L}^\dagger U_{\nu L}$.

Due to the lack of \textit{a priori} knowledge of bulk Dirac mass parameters
and 5D Yukawa couplings, it is not generally possible
to deduce all of their information only from the observed
fermion mass spectrum and mixing angles.
The number of unknown parameters far exceed the number of observations.
One of the best approach is to develop the theory based on a few sound assumptions.
We have the following two natural \textsl{assumptions}:
\begin{enumerate}
    \item For all fermions, 5D Yukawa couplings have a common value $\lambda_5$
of the order of one.
    \item No order-changing by cancellation is allowed
when the multiplication of two mixing matrices explains the observed
mixing matrix.
\end{enumerate}
For \textsl{assumption}-1,
minor differences in $\lm_{5ij}$ are to be absorbed into mixing matrices.
The top quark mass scale is naturally explained by $v \simeq 174\gev$.
Other small SM fermion masses
are generated by controlling $c$'s.
The \textsl{assumption}-2 is consistent with the spirit of no fine-tuning.
When we write the elements of mixing matrices below,
only their order of magnitude does matter.

The \textsl{assumption}-1 leads to the
 following relation for the fermion mass matrix:
\begin{eqnarray}
\label{eq:MfT:Mf}
(M_f^T M_f)_{ij}
= \lm_{5}^2 v^2 F_L(c_{L i}) F_L(c_{L j}) \sum_k F_R(c_{R_k})^2
.
\end{eqnarray}
Since the left-handed up-type ($u_{iL}$ or $\nu_{iL}$)
and down-type ($d_{iL}$ or $e_{iL}$) belong to the same SU(2)$_L$ doublet
and thus have the same $c$,
Eq.\,(\ref{eq:MfT:Mf}) shows the proportionality of
\beq
\label{eq:proportional}
M_u^T M_u\propto M_d^T M_d,\quad M_\nu^T M_\nu \propto M_e^T M_e.
\eeq
Using the relation of
\beq
\label{eq:MTM:UMd2U}
M_f^T M_f = U_{fL}\big( M_f^{\rm (d)} \big)^2 U_{fL}^\dagger ,
\eeq
the proportionality in Eq.\,(\ref{eq:proportional})
helps determine the bulk Dirac mass parameters,
if $U_{fL}$ is known.

\subsection{Quark Sector Mass and Mixing}
\label{subsec:quark}

In the quark sector,
\textsl{assumption}-1 and -2 are enough to fix the model
due to the hierarchical masses and almost diagonal
mixing matrix.
Nine Dirac mass parameters ($c_{Q_i}$, $c_{U_i}$, $c_{D_i}$) are fairly well
determined\,\cite{Huber:2003tu,Chang:2005ya,Agashe:2004ay,Agashe:2004cp}.
The \textsl{assumption}-2 can be easily satisfied if
both $U_{uL}$ and $U_{dL}$ are CKM-type:
The $V^{\rm CKM}= U_{uL}^\dagger U_{dL}$ condition
is naturally satisfied without any fine-tuned cancellation.
We parameterize
\begin{equation}
\left( U_{qL} \right)_{ij} = \kappa_{ij} V^{\rm CKM}_{ij},
 \label{mixing-quark}
\end{equation}
where $\kappa_{ij}$'s are complex parameters of the order of one.
To avoid
order changing during the diagonalization of matrix,
we take $|\kappa_{ij}| \in [1/\sqrt{2},\sqrt{2}]$.

In the simplified Wolfenstein parametrization
with $\lambda \simeq 0.22$, the CKM matrix is
\begin{equation}
V^{\rm CKM}\simeq
\left(\begin{array}{ccc} 1 &\lambda & \lambda^3 \\
\lambda & 1 & \lambda^2 \\ \lambda^3 & \lambda^2 & 1
\end{array} \right). \label{CKM}
\end{equation}
With the observed SM quark mass spectra of
\beq
M_u^{\rm (d)}  \simeq v\  {\rm diag}(\lambda^8, \lambda^{3.5}, 1),
\quad
M_d^{\rm (d)}  \simeq v\  {\rm diag}(\lambda^7, \lambda^5, \lambda^{2.5} ),
\label{quarkmass}
\eeq
we get
\beq
\label{eq:MuMu:MdMd}
U_{uL} (M_u^{\rm (d)})^2 U_{uL}^T
\simeq v^2
\left(\begin{array}{ccc} \lambda^6 &\lambda^5 & \lambda^3 \\
\lambda^5 & \lambda^4 & \lambda^2 \\ \lambda^3 & \lambda^2 & 1
\end{array} \right),
\quad
U_{dL} (M_d^{\rm (d)})^2 U_{dL}^T
\simeq  v^2 \lm^5
\left(\begin{array}{ccc}
\lambda^{6} &\lambda^{5} & \lambda^3 \\
\lambda^{5} & \lambda^4 & \lambda^2 \\
\lambda^3 & \lambda^2 & 1
\end{array} \right).
\eeq
Comparing two matrices in Eq.\,(\ref{eq:MuMu:MdMd}) based on
Eqs.\,(\ref{eq:proportional}) and (\ref{eq:MTM:UMd2U}),
we have
\bea
\label{eq:hierarchy:relation}
F_L(c_{Q_1}) : F_L(c_{Q_2}) : F_L(c_{Q_3}) &\simeq & \lm^3 : \lm^2 : 1,
\\ \no
F_R(c_{A_1}) : F_R(c_{A_2}) : F_R(c_{A_3}) &\simeq & \lm^3 : \lm^2 : 1
,\qquad \hbox{for } A=U,D \,,
\\ \no
F_R(c_{D_1})^2+F_R(c_{D_2})^2+F_R(c_{D_3})^2 &\simeq& \lm^5
\left[
F_R(c_{U_1})^2+F_R(c_{U_2})^2+F_R(c_{U_3})^2
\right]
\,.
\eea

Therefore, the SM quark mixing matrices can be approximated as
\begin{eqnarray}
\label{eq:UqL:UqR}
 \left( U_{qL} \right)_{ij(i\leq j)} \approx
\frac{F_L(c_{Q_i})}{F_L(c_{Q_j})},\quad
\left( U_{qR} \right)_{ij(i\leq j)} \approx
\frac{F_R(c_{A_i})}{F_R(c_{A_j})}.
\label{mixing1}
\end{eqnarray}
The bulk Dirac mass
parameters are determined, as in Ref.\,\cite{Chang:2005ya},
\begin{eqnarray}
\label{eq:c:quark} c_{Q_1} &\simeq &\phantom{-}0.61,
\quad c_{Q_2}  \simeq \phantom{-} 0.56 ,\quad c_{Q_3}
\simeq \phantom{-} 0.3\,^{+0.02}_{-0.04},
\\ \no
c_{D_1} &\simeq& -0.66 ,\quad c_{D_2} \simeq -0.61 ,\quad c_{D_3} \simeq -0.56 \
,
\\ \no
c_{U_1} &\simeq& -0.71 ,\quad c_{U_2} \simeq -0.53 ,\quad 0\lsim c_{U_3} \lsim
0.2.
\end{eqnarray}

\subsection{Lepton Sector Mass and Mixing}
\label{subsec:lepton}

The Pontecorvo--Maki--Nakagawa--Sakata
(PMNS) mixing matrix in the weak charged lepton current is approximately
\beq \label{eq:Upmns}
U_{\rm PMNS}  \simeq
\left(
  \begin{array}{ccc}
    ~0.8~ & ~0.5~ & ~U_{e3}~ \\
    0.4 & 0.6 & 0.7 \\
    0.4 & 0.6 & 0.7 \\
  \end{array}
\right),
\eeq
where the current data constrains $U_{e3} \lsim 0.18 $ at $2\,\sigma$\,\cite{Fogli:2006yq}.
Since $\pmns = U_{e L}^\dagger U_{\nu L}$,
the elements of the $(2,3)$ block
of $U_{eL}$ and $U_{\nu L}$
are of the order of one.
In addition,
the specific form of mass matrix in Eq.\,(\ref{eq:MfT:Mf}) allows
only the normal mass hierarchy for the neutrino masses.
The observed lepton masses are then
\beq
\label{eq:nomal:mass}
M_\nu^{\rm (d)} \simeq m_{\nu_3} \; {\rm diag}(0, \dt, 1),
\quad
M_e^{\rm (d)} \simeq m_{\tau} \; {\rm diag}(\;\dt^4, \dt^{1.5}, 1),
\eeq
where $\delta = \sqrt{ \Dt m_{\rm sol}^2/\Dt m_{\rm atm}^2} \approx 0.173$.
Numerical estimation shows that $\delta \simeq U_{e3}$ in this model \cite{Chang:2005ya}.
Then the condition of $U_{eL}( M_e^{\rm (d)} )^2 U_{eL}^\dagger
\propto U_{\nu L}( M_\nu ^{\rm (d)} )^2 U_{\nu L}^\dagger  $
from Eqs.\,(\ref{eq:proportional}) and (\ref{eq:MTM:UMd2U}) leads to
\beq
\label{eq:13}
\left( U_{eL} \right) _{13}  \simeq  \left( U_{\nu L} \right) _{13}.
\eeq

Using the unitarity condition of mixing matrices,
$U_{\nu L}$ is well constrained as
\beq
\label{eq:UnuL}
U_{\nu L} \sim
\left(
  \begin{array}{ccc}
    1 & 1 &  \dt \\
    1 & 1 & 1 \\
    1 & 1 & 1 \\
  \end{array}
\right).
\eeq
Substituting
$U_{\nu L}$ in $M_\nu^T M_\nu$,
\beq
M_\nu^T M_\nu = U_{\nu L} (M_\nu^{\rm (d)})^2 U_{\nu L}^\dagger
\propto
\left(
  \begin{array}{ccc}
    \dt^2 & \dt & \dt \\
    \dt & 1 & 1 \\
    \dt & 1 & 1 \\
  \end{array}
\right),
\eeq
we have the following relations among $F_L(c_{L_i})$
from the definition in Eq.\,(\ref{eq:MfT:Mf}):
\beq
\label{eq:condition:delta}
\dt \simeq \frac{F_L(c_{L_1})}{F_L(c_{L_2})}
\simeq \frac{F_L(c_{L_1})}{F_L(c_{L_3})}.
\eeq
From the behavior of $F_L(c)$ in Fig.\,\ref{fig:FL},
we have the following hierarchy:
\beq
\label{eq:cL}
 c_{L_2}  \simeq c_{L_3} < c_{L_1} .
\eeq
The relation $U_{e L} (M_e^{\rm (d)})^2 U_{e L}^\dagger \propto M_\nu^T M_\nu$
leaves minor ambiguity in
$U_{eL}$:
\begin{equation}
\label{eq:UeL}
U_{eL}\sim \left(\begin{array}{ccc}
1 & u_{12} & \dt \\
u_{21}  & 1& 1 \\
u_{31}  & 1& 1
    \end{array}\right),
    \qquad \hbox{ for }~
u_{12}  \lsim \dt, \quad u_{21} + u_{31} \simeq \dt.
\end{equation}
The condition of $u_{21} + u_{31} \simeq \dt$
comes from the unitarity of $U_{eL}$, {\it i.e.},
$\left( U_{e L}^\dagger U_{eL} \right)_{13} =0$.

The matrix form of $ U_{eR}$ is well determined due to the hierarchical
charged-lepton masses.
We attribute $F_R(c_{E_i})$ to the source of the hierarchy.
The right-handed lepton mixing matrix should have an approximately symmetric form
of
\begin{eqnarray}
\label{eq:UeR}
\left( U_{eR} \right)_{ij}
\approx\frac{F_R(c_{E_i})}{F_R(c_{E_j})}
\quad \mbox{ for } i\leq j.
\end{eqnarray}

Unlike in the quark sector, the lepton mass spectrum and mixing information
are not enough to fix all the values of the bulk Dirac mass parameters
for the SU(2)$_L$ doublet.
We will resort to
the phenomenological constraint from
the lepton flavor violating decays of $\mu$ and $\tau$
to reduce the ambiguity below.

\section{FCNC through KK gauge bosons}
\label{sec:FCNC}

In this model, the mass eigenstate of the SM fermion is a mixture of
gauge eigenstates as in Eq.\,(\ref{eq:chi-psi-mix}).
Since the 5D gauge interaction is flavor diagonal,
we have FCNC mediated by KK gauge bosons.
We denote $W^{(n)}_{L}$, $W^{(n)}_{R}$ and $B^{(n)}_X$
for the KK gauge fields of \gauge, respectively.
Their 5D gauge couplings ($g_{5L}$, $g_{5R}$ and $g_{5X}$)
are related with the 4D effective couplings through
\begin{eqnarray}
 g&=& g_{4L}=\frac{g_{5L}}{\sqrt{kL}},\nonumber\\
\tilde{g}&=& g_{4R}=\frac{g_{5R}}{\sqrt{kL}}\sim g',\nonumber\\
g_X&=&g_{4X}=\frac{g_{4Y}g_{4R}}{\sqrt{g_{4R}^2-g_{4Y}^2}}\sim g'.
\end{eqnarray}

In terms of gauge eigenstates, the 4D gauge interactions with KK gauge modes
are
\beq
\mathcal{L}_{4D} \supset
g^a_{4D} \sum_{n=1}^\infty
\left(\hat{g}_{L}^\n(c_{i})\, \bar{\psi}_{iL}^{(0)} T^a\gamma^\mu
\psi_{iL}^{(0)}+\hat{g}_{R}^\n(c_{i})\, \bar{\psi}_{iR}^{(0)} T^a\gamma^\mu
\psi_{iR}^{(0)}\right) A^{a\n}_\mu,
\eeq
where $T^a=(T_L,T_R,Y_X)$ for
$A^a=(W_{3L},W_{3R},B_X), Y_X=(B-L)/2$, $g^a_{4D} = g^a_5/\sqrt{kL}$ and
\begin{eqnarray}
\hat{g}^\n_{L}(c_{f_i}) &=& \sqrt{k L} \int \di z k \left[
f_L^{(0)}(z,c_{f_i}) \right]^2 f_A^\n(z)\equiv\hat{g}^\n(c_{f_i}),
\nonumber\\
\hat{g}^\n_{R}(c_{f_i}) &=& \sqrt{k L} \int \di z k \left[
f_R^{(0)}(z,c_{f_i}) \right]^2 f_A^\n(z)=\hat{g}^\n(-c_{f_i})\,.
\end{eqnarray}

\begin{figure}
 \includegraphics[scale=0.6]{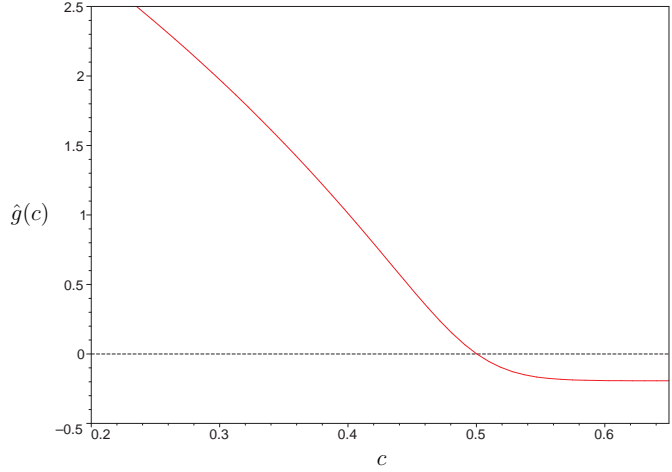}
\caption{Plot of $\hat{g}^{(1)}(c)$ for bulk mass $c$}
\label{fig:ghat}
\end{figure}

For later discussions we plot
$\hat{g}^\n(c)$
as a function of $c$
in Fig.\,\ref{fig:ghat}.
Note that $\hat{g}(c)$
vanishes at $c=1/2$:
\beq
\hat{g}^\n \left( c= \frac{1}{2} \right) = 0.
\eeq
Another interesting feature is that the value of $\hat{g}^\n(c)$
converges into $ -0.2$ for $c \gsim 0.55$.
As $c$ becomes less than 1/2,
the value of $\hat{g}^\n(c)$ increases rather sharply.

Considering the $B\to K^*l^+l^-$ process, we focus
on the mixing among the SM down-type quarks $d_i$
and the SM charged leptons $e_i$,
mediated by the
$n$-th neutral KK gauge bosons in this model:
\begin{eqnarray}
\label{eq:FV:Lg}
\mathcal{L}_{4D} \supset
-\frac{1}{2}\sum_{i,j,n} &\biggl[ &
g
\left(K^\n_{Qij}\,\bar{d}_{iL}  \gamma^\mu d_{jL}
+ K^\n_{Lij}\,\bar{e}_{iL}  \gamma^\mu e_{jL}
\right) W^\n_{3L\mu}
\\
&& \hspace*{-0.5cm} +\;
\tilde{g}
\left(K^\n_{Dij}\,\bar{d}_{iR}  \gamma^\mu d_{jR}+ K^\n_{Eij}\,
\bar{e}_{iR}  \gamma^\mu e_{jR}
\right) W^\n_{3R\mu} \nonumber\\\nonumber
&&\hspace*{-0.5cm} -\;
g_X
\left( K^\n_{Qij}\,\bar{d}_{iL}  \gamma^\mu d_{jL}
-  K^\n_{Lij}\,\bar{e}_{iL}  \gamma^\mu e_{jL}
+
 K^\n_{Dij}\,\bar{d}_{iR}  \gamma^\mu d_{jR}-
 K^\n_{Eij}\,\bar{e}_{iR}  \gamma^\mu e_{jR}
\right) B^\n_{X\mu} \biggl],
\end{eqnarray}
where $i,j$ are the generation indices ($i,j=1,2,3$),
and
\begin{eqnarray} \label{eq:Kdef}
K^{\n}_{Qij}&=& \sum_{k=1}^3 \big( U_{dL}^\dagger \big)_{ik}
\,\hat{g}^{\n}(c_{Q_k}) \left( U_{dL} \right)_{kj} \,,
\nonumber\\
K^{\n}_{Dij}&=& \sum_{k=1}^3 \big( U_{dR}^\dagger \big)_{ik}
\,\hat{g}^{\n}(-c_{D_k}) \left( U_{dR} \right)_{kj} \,,
\nonumber\\
K^{\n}_{Lij}&=& \sum_{k=1}^3 \big( U_{eL}^\dagger \big)_{ik}
\,\hat{g}^{\n}(c_{L_k}) \left( U_{eL} \right)_{kj} \,,
\nonumber\\
K^{\n}_{Eij}&=& \sum_{k=1}^3 \big( U_{eR}^\dagger \big)_{ik}
\,\hat{g}^{\n}(-c_{E_k}) \left( U_{eR} \right)_{kj} \,.
\end{eqnarray}

\section{Flavor Violating Process}
\label{sec:FV}

\subsection{Lepton Flavor Violations}

In this model, the flavor-violating interactions in Eq.\,(\ref{eq:FV:Lg})
generates the lepton-flavor-violating decay of $l \to l' l'' l'''$
at tree level, which is mediated by KK gauge bosons.
Radiative lepton-violating processes such as $\mu \to e \gamma$
does not happen at tree level.
With negligible SM contributions,
the bulk-RS effects become dominant for
$\tau\rightarrow 3e$ and $\tau\rightarrow 3\mu$
with the following experimental bound\,\cite{Yao:2006px}:
\bea
 \frac{\Gamma(\tau\rightarrow 3e)}{\Gamma(\mu\rightarrow  e\nu_\mu\bar\nu_e)}
&\simeq&
(K^{(1)}_{L11}K^{(1)}_{L12})^2\left(\frac{m_Z}{M^{(1)}_A}\right)^4
\lsim  1.0 \times 10^{-12} , \label{eq:mu:3e}
\\
 \frac{\Gamma(\tau\rightarrow 3\mu)}{\Gamma(\tau\rightarrow
\mu\nu_\tau\bar\nu_\mu)}
&\simeq&
(K^{(1)}_{L22}K^{(1)}_{L23})^2\left(\frac{m_Z}{M^{(1)}_A}\right)^4
\lsim 10^{-6} , \label{eq:tau:3mu}
\eea
where
$m_Z$ is the SM $Z$ boson mass.
Here we consider only the major contributions from the lightest KK gauge boson
since the bulk-RS effect is suppressed by
the forth power of $M_A^\n$.

If $M^{(1)}_A\leq 3$ TeV, Eqs.\,(\ref{eq:mu:3e}) and (\ref{eq:tau:3mu})
constrain
\bea
c_{L_2}\simeq c_{L_3}\simeq 0.5 \label{eq:clr}.
\eea
We justify it for the case of $M^{(1)}_A= 3$ TeV as follows.
Substituting $U_{eL}$ in Eq.\,(\ref{eq:UeL})
into $K_{Lij}$ in Eq.\,(\ref{eq:Kdef}),
the  $\tau \to 3 \mu$ constraint in Eq.\,(\ref{eq:tau:3mu}) becomes
\begin{equation}
K^{(1)}_{L22}K^{(1)}_{L23}\simeq (\hat{g}_2+\hat{g}_3)^2
\simeq (2 \hat{g}_2)^2
< 1 ,\label{eq:cond1}
\end{equation}
where $\hat{g}_i \equiv \hat{g}^{(1)} (c_{L_i})$,
and the second equality comes from Eq.\,(\ref{eq:condition:delta}).
We have also used $\hat{g}_{1,2,3} \lsim \mathcal{O}(1)$,
and $u_{12,21,31} \lsim \dt$. From the functional behavior of $\hat{g}(c)$
in Fig.\,\ref{fig:ghat}, we have $c_{L_2}\simeq c_{L_3} > 0.45$.
This mild condition on $c_{L_2}$ and $c_{L_3}$
constrains $c_{L1}>0.55$ and thus
 $| \hat{g}_1 | \simeq 0.2 $,
 as can be seen from Eq.\,(\ref{eq:condition:delta}) and Figs.\,\ref{fig:FL}
 and \ref{fig:ghat}.
More constraint on $c_{L_{2,3}}$ comes from $\mu \to 3 e$:
\bea
K^{(1)}_{L11}K^{(1)}_{L12} &\simeq&
 \,
\hat{g}_1
\left\{ \,
 \hat{g}_1 u_{12} + \hat{g}_2 (u_{21} +u_{31})
\right\}
\\ \no
&\simeq&
\hat{g}_1
\left\{ \,
 \hat{g}_1 u_{12} +  \hat{g}_2 \dt
\right\}
\lsim  10^{-3}\quad \hbox{for } M_A^{(1)} \simeq 3\tev, \label{eq:mutoe2}
\eea
where the second equality is from Eq.\,(\ref{eq:UeL}).
The saturating value  $|\hat{g}_1 |\approx 0.2$ suppresses
the $u_{12}$ element of $U_{eL}$ to be very small,
$u_{12} < \dt^2$.
In addition,
$ \hat{g}_1 \hat{g}_2 \dt \lsim 10^{-3}$ condition requires
$\hat{g}_2 <0.05$.
It strictly constrains such that
$ \left| c_{L_2} -0.5 \right| \leq 0.004$. For $M^{(1)}_A=2$ TeV,
the bound becomes
even more strict:
$ \left| c_{L_2} -0.5 \right| \leq 0.002$.

As a natural solution
for the lepton bulk mass parameters allowed by the current
lepton-flavor violating processes,
we choose
\bea
\label{eq:c:lepton}
c_{L_1}&\simeq& \phantom{-} 0.59,\quad c_{L_2} \simeq \phantom{-}0.5 ,
\quad c_{L_3} \simeq \phantom{-}0.5,\nonumber\\
c_{E_1}&\simeq& -0.74,  \quad c_{E_2} \simeq - 0.65 , \quad c_{E_3} \simeq  -0.55.
\eea

\subsection{Effects on $B \to K^* l^+ l^-$}

The FCNC decay $B \to K^* l^+ l^-$ has been observed
with the branching ratio of the order of $10^{-6}$\,\cite{BKll:exp},
as well as the forward-backward asymmetry\,\cite{BKll:Afb}.
The total transition amplitude for $b\rightarrow s l_i^+l_j^-$ can be written as
\begin{equation}
 {\cal M}={\cal M}_{\rm SM} +{\cal M}_{\rm new} \,.
\end{equation}
For the SM results, we
refer to Ref.\,\cite{Aliev:1999gp,BKllSM}.
For new physics contributions,
we adopt the parametrization in Ref.\,\cite{Aliev:1999gp},
\begin{eqnarray}
 {\cal M}_{\rm new} = \frac{G_F\alpha}{\sqrt{2}\pi}V_{tb}V_{ts}^*
\biggl[ && C_{LL} (\bar{s}_L\gamma^\mu b_L)( \bar{l}_L\gamma^\mu l_L) +
C_{LR} (\bar{s}_L\gamma^\mu b_L)( \bar{l}_R\gamma^\mu l_R)\nonumber\\
&& \!\!\!\! + \,
C_{RL} (\bar{s}_R\gamma^\mu b_R )(\bar{l}_L\gamma^\mu l_L) +
C_{RR} (\bar{s}_R\gamma^\mu b_R)( \bar{l}_R\gamma^\mu l_R )\biggl]\, .
\end{eqnarray}
Note that other new physics parameters ({\it i.e.}, $C_{LRLR}$)
vanish in this model.

The RS contributions can be written as
\begin{eqnarray}
 {\cal M}_{\rm RS}
\simeq \sum_{n=1}^\infty
\frac{1}{4M_A^{(n)2}}
\biggl[ && \left(g^2K^{\n}_{Q23}K^{\n}_{Lii}-
 g_{X}^2 K^{\n}_{Q23} K^{\n}_{Lii}\right)
 (\bar{s}_L\gamma^\mu b_L)( \bar{l}_{iL}\gamma^\mu l_{iL})
\\
&&-
 g_{X}^2  K^{\n}_{Q23} K^{\n}_{Eii}(\bar{s}_L\gamma^\mu b_L)
 ( \bar{l}_{iR}\gamma^\mu l_{iR})
\no
\\
&&
-
g_{X}^2K^{\n}_{D23} K^{\n}_{Lii} (\bar{s}_R\gamma^\mu b_R )
(\bar{l}_{iL}\gamma^\mu l_{iL})
\nonumber\\
&& +
\left(\tilde{g}^2K^{\n}_{D23}K^{\n}_{Eii}-
 g_{X}^2 K^{\n}_{D23} K^{\n}_{Eii}\right) (\bar{s}_R\gamma^\mu b_R)
 ( \bar{l}_{iR}\gamma^\mu l_{iR}) \biggl].\nonumber
\end{eqnarray}
Since physical observables are strongly suppressed by $M_A^\n$,
we consider only the first KK mode effect
and we omit the KK mode number notation $(n)$ in the rest of this section.

The preferred $c_Q$'s in Eq.\,(\ref{eq:c:quark}) and the CKM-type matrices $U_{qL}$
and $U_{qR}$ in Eq.\,(\ref{eq:UqL:UqR}) simplify
$K_{Q23}$ and $K_{D23}$ as
\bea
\label{eq:K23}
K_{Q23}&\simeq& \big( U_{qL} \big)_{23}\big( U_{qL} \big)_{33} \hat{g}(c_{Q_3})
\equiv \kp_Q^2 \hat{g}(c_{Q_3}) V_{tb} V^*_{ts},
 \\
\label{eq:KD23}
K_{D23}&\simeq&
\biggl[ \big( U_{dR} \big)_{22}\big( U_{dR} \big)_{32} \hat{g}(c_{D_2})
 +\big( U_{dR} \big)_{23}\big( U_{dR} \big)_{33} \hat{g}(c_{D_3})
\biggl]
\\ \no
&
\equiv
&
2 \, \kp_D^2  \hat{g}(c_{D_3})
 V_{tb} V^*_{ts},
\eea
where we have used $\hat{g}(c_{Q_3})\gg \hat{g}( c_{Q_{1,2}})$
and  $\hat{g} (c_{D_2}) \approx \hat{g} (c_{D_3})$.
New physics parameters $C_{XX'}$ ($X,X' =L,R$) are
\bea
\label{eq:CXX'}
C_{LL}
&\simeq&
 \left( \frac{\tilde{G}}{M_A} \right)^2(g^2-g_{X}^2)
 \, \kp_Q^2 \, \hat{g}(c_{Q_3}) K_{Lij},
\\ \no
C_{LR} &\simeq&
 \left( \frac{\tilde{G}}{M_A} \right)^2
g_{X}^2
\kp_Q^2 \hat{g}(c_{Q_3})  K_{Eij} , \\ \no
C_{RL} &\simeq&
2 \left( \frac{\tilde{G}}{M_A} \right)^2
 \,g_{X}^2 \kp_{D}^2  \hat{g}(c_{D_3})
 K_{Lij} ,
\\ \no
C_{RR} &\simeq&
2 \left( \frac{\tilde{G}}{M_A} \right)^2 (\tilde{g}^2-g_{X}^2)
 \kp_{D}^2  \hat{g}(c_{D_3})
K_{Eij} \, ,
\eea
where $
 \tilde{G} = \big( {\pi}/{2\sqrt{2} G_F \alpha} \big)^{1/2} \approx 3.5\tev$.

\begin{table}
\caption{\label{tab:CLL's}The values of $C_{LL}$, $C_{RL}$,
$C_{LR}$, and $C_{RR}$ for $b \to s l_i^+ l_j^-$.
We set $M_A^{(1)}=2\tev$, $\kp_{Q,D} =1$, and $\dt=0.15$.
 }
\begin{ruledtabular}
\begin{tabular}{cllllll}
 & $\phantom{-}e^+ e^-$ &$\phantom{-}e^+ \mu^-$ &$\phantom{-}e^+ \tau^-$
    &$\phantom{-}\mu^+ \mu^-$ & $\phantom{-}\mu^+ \tau^-$ & $\phantom{-}\tau^+ \tau^-$ \\
\hline
$C_{LL}$& $ -0.3$ &$\pm\, 7 \times 10^{-3}$ & $\pm\, 0.05$
    &$- 7\times 10^{-3}$ & $\pm\, 6\times 10^{-3}$ & $-4\times 10^{-5}$\\
$C_{RL}$ & $\phantom{-}0.02$ & $\pm\, 5 \times 10^{-4}$ & $\pm\, 4\times 10^{-3}$
    & $\phantom{-} 6\times 10^{-4}$ & $\pm\, 5\times 10^{-4}$ &
    $\phantom{-}3 \times 10^{-6}$\\
$C_{LR}$ & $-0.1$& $\pm\,0.01$ & $\pm\,10^{-3}$ & $-0.1$ & $\pm\,0.01$ & $-0.1$ \\
$C_{RR}$  & $\phantom{-}0.03$ & $\pm\, 3\times 10^{-3}$ &$\pm\,2\times 10^{-4}$ &
$\phantom{-}0.03$
    & $\pm\,3\times 10^{-3}$ & $\phantom{-}0.02$ \\
\end{tabular}
\end{ruledtabular}
\end{table}

In Table \ref{tab:CLL's}, we present the values  of
$C_{LL}$, $C_{RL}$,
$C_{LR}$, and $C_{RR}$ for $b \to s l_i^+ l_j^-$.
For representative purpose, we set $M_A^{(1)}=2\tev$, $\kp_{Q,D} =1$, $\dt=0.15$,
and use central values of $c$'s in Eqs.\,(\ref{eq:c:quark}) and (\ref{eq:c:lepton}).
The values of $C_{XX'}$
can be understood from $\hat{g}$:
\bea
\hat{g}(c_{Q_3}) &\simeq& 2.0, \quad
\hat{g}(c_{L_2})\simeq \hat{g}(c_{L_3})=0,
\\ \no
\hat{g}(c_{D_3}) &\simeq&  \hat{g}(c_{L_1})\simeq \hat{g}(-c_{E_1})
\simeq  \hat{g}(-c_{E_2}) \simeq  \hat{g}(-c_{E_3}) \simeq -0.2.
\eea
Brief comments on the sign of $C_{XX'}$ are in order here.
The negative  signs of $C_{LL}$ and $ C_{LR}$
are due to positive $\hat{g}(c_{Q_3})$,
and negative $ \hat{g}(c_{L_1})$
and $ \hat{g}(-c_{E_1})$ which dominantly contribute to
$K_{Lij}$ and $K_{Eij}$, respectively.
The sign of $C_{XX'}$ for off-diagonal decays such as $B \to K^* l^+ l^{\prime -}$
is not determined since we could fix only the magnitude of
elements of mixing matrices.
In the magnitudes, only the  $C_{XX'}$'s for $b \to s\, e^+ e^-$
are substantial.
$C_{XX'}$'s for decays involving $\mu^\pm$ or $\tau^\pm$
are quite suppressed,
since $\hat{g}(c_{L_{2}})
\simeq \hat{g}(c_{L_3}) \ll 1$.
Among $C_{XX'}$'s for $b \to s\,e^+ e^-$, $C_{LX}$ is larger than $C_{RX}$
since $\hat{g}(c_{Q_3})$ is much larger than $\hat{g}(c_{D_3})$.

\begin{figure}[t!]
 \includegraphics[scale=0.8]{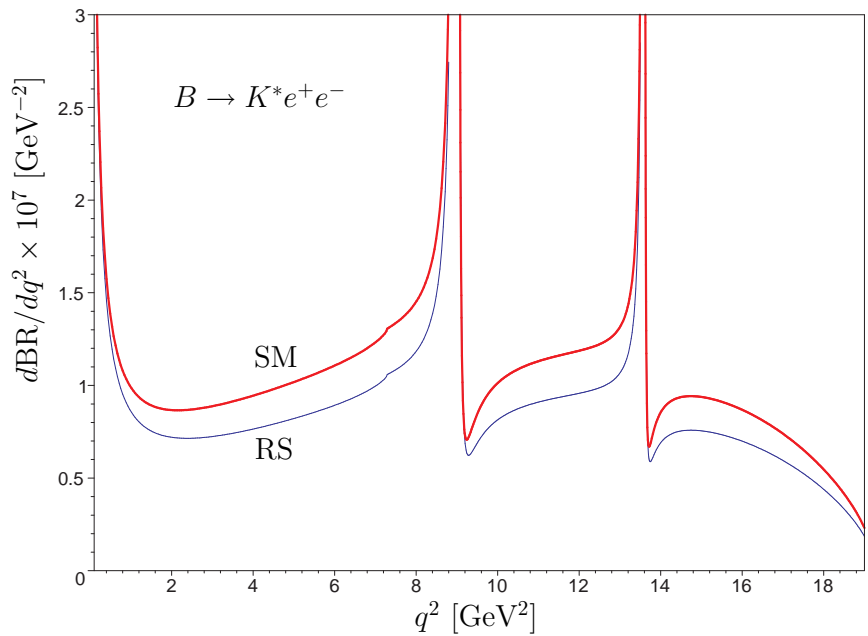}
\caption{$d {\rm BR}/d q^2$ as a function of $q^2$ for $B \to
K^*e^+ e^-$.
The thick (red) line is the SM result,
and the thin (blue) line is for the bulk RS model with $\kp=\sqrt{2}$
and $M_A^{(1)}=2\tev$.} \label{fig:dBRds}
\end{figure}

In Fig.\,\ref{fig:dBRds}, we present
the differential branching ratio $d {\rm BR}/d q^2$ as a function of $q^2$ for $B \to
K^*e^+ e^-$.
We use the following
values for the Wilson coefficients of the SM:
\beq
C_9^{\rm NDR}=4.153,~~~C_{10}=-4.546,~~~C_7=-0.311,
\eeq
which correspond to the next-to-leading QCD corrections \cite{R23,R24}.
The renormalization scale $\mu$ and
the top quark mass are set to be
\beq
\mu=m_b=4.8~{\rm GeV},~~~m_t=175~{\rm GeV}.
\eeq
We follow Refs.\,\cite{longdistance} in taking into account
the long--distance effects of the charmonium states.
For the form factors, we have used the
light-cone QCD sum-rule method predictions\,\cite{formfactor}.
Throughout numerical analysis, we used
the central values of the input parameters,
and do not consider the theoretical uncertainty in the calculation of form factors.
In Fig.\,\ref{fig:dBRds},
the thick (red) line is the SM result,
and the thin (blue) line is for the bulk RS model.
We have used
the allowed maximum value of $C_{XX'}$'s
with $\kp (=\kp_Q =\kp_D)=\sqrt{2}$ and $M_A^{(1)}=2\tev$.
As discussed in Ref.\,\cite{Aliev:1999gp},
this BR distribution is most sensitive to $C_{LL}$.
Since our $C_{LL}$ for $B \to
K^*e^+ e^-$ is negative, the result in this model is less than
in the SM.
The reduction can be maximally about 20\% at some points.
Unfortunately, the theoretical uncertainty of the form factors
are known to be about 15\%\,\cite{formfactor}.
It would be quite challenging for experiments
to probe this new physics effect from the BR distribution.

\begin{figure}
 \includegraphics[scale=0.8]{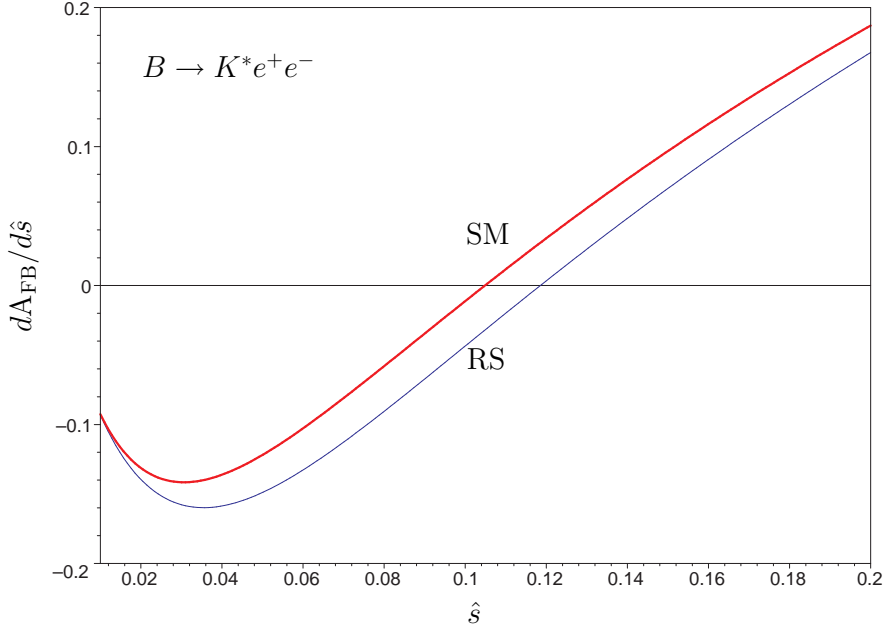}
\caption{$d {\rm A_{FB}}/d \hat{s}$ as a function of $\hat{s}$ for
$B \to K^*e^+ e^-$.
The thick (red) line is the SM result,
and the thin (blue) line is the new physics result
with $\kp=\sqrt{2}$ and $M_A^{(1)}=2\tev$.} \label{fig:dAdshat}
\end{figure}
One sensitive observable to new physics
is known to be the zero value position
of the forward-backward asymmetry, {\it i.e.},
$\afb (\hat{s}_0)=0$.
The forward-backward asymmetry $\afb(\shat)$ is defined by
\beq
\dfrac{d}{d \shat} \afb (\shat) =
\dfrac{\int_0^1  d z \dfrac{d \Gm}{d \shat \, d z} - \int_{-1}^0 \dfrac{d \Gm}{d \shat\, d z}}
{\int_0^1 d z \dfrac{d \Gm}{d \shat \, d z} + \int_{-1}^0 \dfrac{d \Gm}{d \shat \,d z}},
\eeq
where $\shat=q^2/m_B^2$, $z=\cos\theta$, and $\theta$ is the angle
between $K^*$ and $l^-$.
In the large energy expansion theory,
it has been shown that $\hat{s}_0$
has no hadronic uncertainty;
it is determined simply by the short-distance Wilson coefficients $C_9^{\rm eff}$
and $C_7^{\rm eff}$\,\cite{s0}.
In Fig.\,\ref{fig:dAdshat}, we show the $\afb(\shat)$
as a function of $\shat$.
The thick (red) line is the SM result,
and the thin (blue) line is the new physics result
with $\kp=\sqrt{2}$ and $M_A^{(1)}=2\tev$.
The zero value position of $\afb$ in the SM model is consistent with
other results\,\cite{Arda:2002iq}.
In our new model,
$\hat{s}_0$ shifts to the positive direction:
$\hat{s}_0$ can increase maximally about 18\%.
Experimental sensitivity
is expected to reach this difference in  near future.

\begin{figure}[t!]
 \includegraphics[scale=0.8]{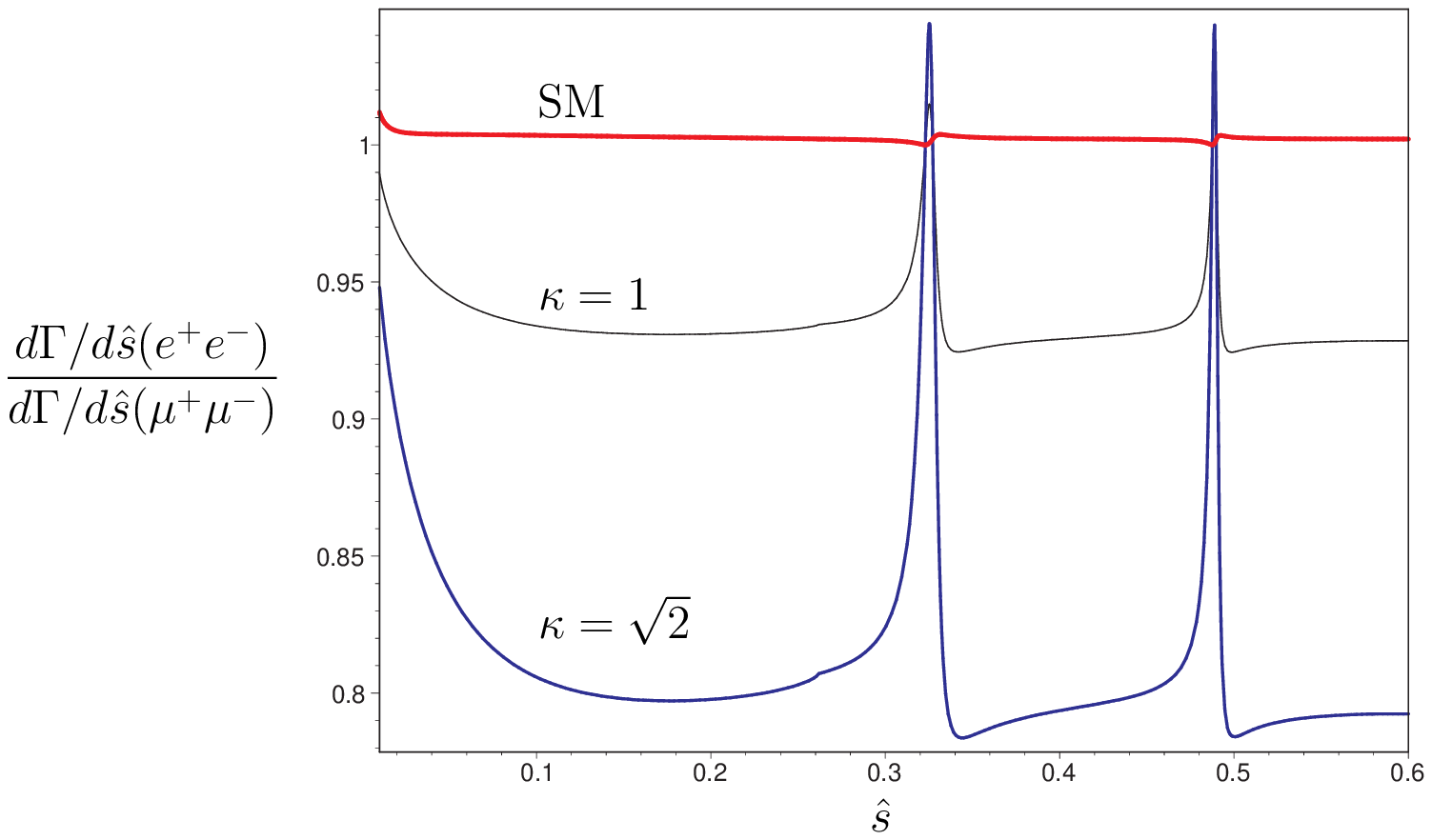}
\caption{$\dfrac{d \Gm }{d \hat{s}}(B \to K^* e^+ e^-) \biggl{/}
\dfrac{d \Gm }{d \hat{s} }(B \to K^* \mu^+ \mu^-)
$ as a function of $\hat{s}$ in the SM and the custodial bulk RS with
$\kp=1,\sqrt{2}$.
we set $M_A^{(1)}=2\tev$.
} \label{fig:dGmdshat_ratio}
\end{figure}

Now we present new phenomenological signatures exclusively for this model.
One of the most unique features
is that only the $B \to K^* e^+ e^-$ decay has
sizable new physics effect while others have
negligible effects.
Therefore, we consider
the ratio of differential decay rate of
$B \to K^* e^+ e^-$ to that of  $B \to K^* \mu^+ \mu^-$.
This ratio was proposed as an efficient observable to test the SM\,\cite{Hiller}.
In Fig.\,\ref{fig:dGmdshat_ratio},
we show the ratios as a function of $\shat$
in the SM and the bulk RS model.
The thick (red) line is for the SM result,
the thin (black) line for the RS result with $\kp=1$,
and the normal (blue) line for the RS result with $\kp=\sqrt{2}$.
In the most range of $\shat$,
the RS result is far below the SM one.
For maximally allowed value of $C_{XX'}$ with $\kp=\sqrt{2}$,
the deviation from the SM result can be about 20\% for sizable range of $\shat$.
Even for moderate values of $C_{XX'}$ with $\kp=1$,
the deviation reaches up to 7\%.
Moreover, as taking the \emph{ratio} of differential decay rates,
most of the hadronic uncertainty in the calculation of form factors
disappears.
This can be a quite clean signal for experiments.

\begin{figure}
 \includegraphics[scale=0.9]{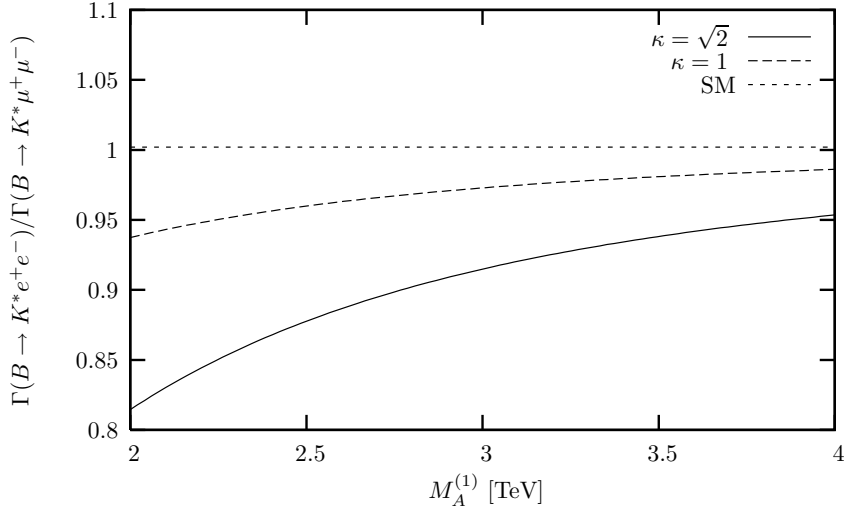}
\caption{$\dfrac{ \Gm(B \to K^* e^+ e^-)}{\Gm(B \to K^* \mu^+ \mu^-) }
$ as a function of $M_A^{(1)}$ in the SM and the custodial
bulk RS with $\kp=1,\sqrt{2}$.
Here $\Gm$ is partially integrated $d \Gm /d \shat$ for
$\shat \in [0.1,\;4\,{m}_c^2/m_B^2]$.
}
\label{fig:Gm_ratio}
\end{figure}

In order to see the dependence of new physics effect
on $M_A^{(1)}$,
we present
the ratio of two partially integrated decay rates
for $B \to K^* e^+ e^-$ and $B \to K^* \mu^+ \mu^-$.
From the profile in Fig.\,\ref{fig:Gm_ratio} as a function of $\shat$,
we choose the integration range of
$\shat \in [\, 0.1,\;4{m}_c^2/m_B^2]$
with $m_c$ being the charm quark mass.
The dotted line is for the SM result,
the dashed line for the RS result with $\kp=1$,
and the solid line for the RS result with $\kp=\sqrt{2}$.
If $C_{XX'}$'s have allowed maximum values ($\kp=\sqrt{2}$),
the RS result with $M_A^{(1)}=2\tev$ shows about 18\% deviation from the SM result,
and that even with $M_A^{(1)}=4\tev$ shows about 4.5\%.
If $C_{XX'}$'s have medium values ($\kp=1$),
the RS result with $M_A^{(1)}=2\tev$ shows about 6\% deviation from the SM result,
and that with $M_A^{(1)}=4\tev$ shows about 2\%.
Since the ratio does not suffer from the hadronic uncertainty of the form factors,
this difference will be within experimental sensitivity in  near future.

\section{Conclusions}
\label{sec:conclusions}

The custodial Randall-Sundrum model is a warped 5D model with all the SM fields in
the bulk.
Only the Higgs boson field is confined on the TeV brane,
which generates masses for the SM particles.
The troublesome EWPD constraint is overcome
by SU(2) custodial  symmetry
induced from
AdS$_5$/CFT feature of bulk gauge symmetry of \gauge.
We focused on new FCNC phenomena which occur due to the misalignment between
the gauge couplings and the 5D Yukawa interaction.
We have the vertex of $f$-$f'$-$A^{(n)}$,
where $f^{(\prime)}$ is a SM fermion and $A^{(n)}$ is a Kaluza-Klein mode of a neutral gauge boson.
At tree level, we have non-SM FCNC involving four external SM fermions,
mediated by KK neutral gauge bosons.

The $f$-$f'$-$A^{(n)}$ vertex depends on two kinds of model parameters,
the 5D Yukawa couplings  and the bulk Dirac mass parameters.
They also determine the SM fermion mass spectrum and mixing angles.
Based on two natural assumptions of universal 5D Yukawa coupling and
no-cancellation in explaining the observed SM fermion mixing matrices,
we have obtained all the information on $c$'s as well as mixing angles.

In the custodial bulk RS model with very specified fermion structure,
we study  FCNC process of $B \to K^* l^+ l^{\prime -}$.
New physics effect is parameterized in the helicity amplitude
as $C_{XX'} (\bar{s}_X\gamma^\mu b_X)( \bar{l}_{X'}\gamma^\mu l_{X'})$,
where $X,X'=L,R$.
If $C_{XX'}$'s have maximally allowed values,
the differential decay rate of $B \to K^* e^+ e^-$
deviates from the SM result as much as about 20\% at some $q^2$.
Unfortunately, the hadronic uncertainty in the form factors
is large enough to sweep away this new effect.
Instead, the zero value point of the forward-backward asymmetry, $\shat_0$, is
known to be
quite insensitive to the hadronic uncertainty.
In the maximal case, the deviation of $\shat_0$ from the SM value is about 5\%,
which is expected to be probed in  near future.

We have also found  the following characteristic features:
\begin{itemize}
    \item The best chance to observe the custodial bulk RS model effect is
through $b \to s \, e^+ e^-$ due to the suppressed couplings of
$\mu^+$-$\mu^-$-$Z^\n$
and $\tau^+$-$\tau^-$-$Z^\n$.
And $C_{LL}$ is dominant, and $C_{LR}$ is the second dominant.
    \item Two other decays of $b \to s \mu^+ \mu^-$
and $b \to s \tau^+ \tau^-$
have dominant vertex of $C_{LR}$.
Unfortunately, their magnitudes are too small for experiments to probe in  near future.
    \item Other non-diagonal decay modes of $b \to s l^+_i l^-_j (i \neq j)$
are quite suppressed in this model.
\end{itemize}
Based on these observations,
we suggested new phenomenological signatures to probe the custodial bulk-RS model.
The first one is the ratio of two differential decay rates
of $B \to K^* e^+ e^-$ and $B \to K^* \mu^+ \mu^-$.
Upon taking the ratio,
the hadronic uncertainty in the calculation of form factors becomes negligible.
Since two decay modes in the SM have almost the same decay rates
with slight kinematic difference from the lepton masses,
dominant new physics effects only for $B \to K^* e^+ e^-$
leads to sizable deviation from the SM result.
For $M_A^{(1)}=2 ~(4) \tev$,
the deviation can reach about 20\% (7\%).
We also showed the ratio of partially integrated decay rates,
which shows also about 10-20\% deviation from the SM results.
This deviation is expected to be observed in near future.

\acknowledgments
\noindent The work of C.S.K. was supported
in part by  CHEP-SRC Program, and
in part by the Korea Research Foundation Grant funded
by the Korean Government (MOEHRD) No. KRF-2005-070-C00030.
The work of S.C. and J.S. was supported
by the Korea Research Foundation Grant. (KRF-2005-070-C00030).

\end{document}